\documentstyle[12pt]{article}
\begin{document}

\def\a{\alpha}
\def\b{\beta}
\def\ch{\chi}
\def\d{\delta}
\def\e{\epsilon}
\def\f{\phi}
\def\g{\gamma}
\def\h{\eta}
\def\i{\iota}
\def\j{\psi}
\def\k{\kappa}
\def\l{\lambda}
\def\m{\mu}
\def\n{\nu}
\def\o{\omega}
\def\p{\pi}
\def\q{\theta}
\def\r{\rho}
\def\s{\sigma}
\def\t{\tau}
\def\u{\upsilon}
\def\x{\xi}
\def\z{\zeta}
\def\D{\Delta}
\def\F{\Phi}
\def\G{\Gamma}
\def\J{\Psi}
\def\L{\Lambda}
\def\O{\Omega}
\def\P{\Pi}
\def\S{\Sigma}
\def\U{\Upsilon}
\def\X{\Xi}
\def\T{\Theta}

\def\Ab{\bar{A}}
\def\gi{g^{-1}}
\def\li{{ 1 \over \l } }
\def\lb{\l^{*}}
\def\zb{\bar{z}}
\def\ub{u^{*}}
\def\Tb{\bar{T}}
 \def\pp {\partial }
\def\pb {\bar{\partial }}
\def\be{\begin{equation}}
\def\ee{\end{equation}}
\def\ben{\begin{eqnarray}}
\def\een{\end{eqnarray}}

\addtolength{\topmargin}{-0.8in}
\addtolength{\textheight}{1in}
\hsize=16.5truecm
\hoffset=-.5in
\baselineskip=7mm

\thispagestyle{empty}
\begin{flushright} \ Feb. \ 1994\\
KHTP-94-01/ SNUCTP 94-09\\
\end{flushright}

\begin{center}
 {\large\bf  Deformed Coset Models From Gauged WZW Actions
 }\\[.1in]
\vglue .7in
Q-Han Park\footnote{ E-mail address; qpark@nms.kyunghee.ac.kr }
\vglue .5in
{\it Department of Physics, Kyunghee University\\
Seoul, 130-701, Korea}
\\[.7in]
{\bf ABSTRACT}\\[.2in]
\end{center}
\vglue .2in
A general Lagrangian formulation of  integrably deformed  G/H-coset
models is given. We  consider  the G/H-coset model in terms of
the gauged Wess-Zumino-Witten action and obtain an integrable deformation
 by adding a potential energy  term $Tr(gTg^{-1}\Tb )$,
where algebra elements $T, \Tb $ belong to the center of the algebra {\bf h}
associated with the subgroup H. We show that the classical equation of motion
of the  deformed coset model  can  be identified with  the integrability
condition of  certain linear equations  which makes the use of  the
inverse scattering method possible.
Using the linear equation,  we give a systematic way to construct infinitely
many  conserved currents as well as soliton solutions. In the case of the
parafermionic  SU(2)/U(1)-coset model, we derive $n$-solitons and conserved
currents explicitly.

\newpage

Integrable field theories in two dimensions (2-d IFT) have been
successfully developed in the past decades.  Recently,  Zamolodchikov has
shown that there exist deformations of conformal field theories (2-d CFT) with
relevant operators which preserve integrability[1].
In the case of certain rational conformal field theories, these deformations
have been explained in terms of the affine extension of the Toda field
theory[2][3]. However, a general Lagrangian framework for integrably deformed
2-d CFT's has not been known so far.

The purpose of this letter is to provide a  Lagrangian formulation
of general G/H-coset models and their integrable deformations in the context of
the gauged Wess-Zumino-Witten (WZW) model[4][5]. We show that an integrable
deformation
of general G/H-coset models is possible when the gauged WZW action for
the G/H-coset model is added by  a potential energy term $Tr(gTg^{-1}\Tb )$,
where
algebra elements $T, \Tb $ belong to the center of the algebra {\bf h}
associated with the subgroup H.
In the case of SU(2)/U(1), this reduces to the Lagrangian, given recently by
Bakas[6], of the parafermion model deformed by the energy operator
$\Phi_{1,3}$.
The main observation of this work is that the classical equation of motion of
the deformed coset model takes the form of a zero curvature which can be
identified with  the integrability condition  of the associated linear
equations with  a spectral parameter. This allows us to apply the inverse
scattering method to the problem and using this method, we give a systematic
way to construct infinitely many conserved currents and $n$-soliton solutions.
As an example, we construct explicitly conserved currents and $n$-soliton
solutions of the deformed parafermionic SU(2)/U(1)-coset model.

We first recall that a Lagrangian of the G/H-coset model is given
 in terms of the gauged  WZW functional[4][5], which in light-cone variables
 is
\be
I(g,A, \Ab ) = I_{WZW}(g) +
{1 \over 2\pi }\int \mbox {Tr} (- A\pb g \gi + \Ab \gi \pp g
 + Ag\Ab \gi - A\Ab )
\ee
where $I_{WZW}(g)$ is the usual WZW action [5] for a map $g  :  M \rightarrow
$G
 on two-dimensional Minkowski space $M$.  The connection $A, \Ab$ gauge the
anomaly
free subgroup H of G. In this letter, we take the diagonal
embedding of H in $G_{L} \times G_{R}$, where $G_{L}$ and $G_{R}$ denote left
and right group actions by multiplication $(g \rightarrow g_{L}gg_{R}^{-1})$,
so that Eq.(1) becomes invariant under the vector gauge transformation $(g
\rightarrow hgh^{-1}$ with $h :  M \rightarrow $H).
The key observation to be made is that the equation of motion of
the gauged WZW action takes the form of a zero curvature,
\be
\d_{g}I(g,A,\Ab ) = -{1 \over 2\pi }\int \mbox{Tr}
[\  \pp + \gi \pp g + \gi A g \ , \ \pb + \Ab \ ]\gi \d g  = 0
\ee
which, together with the constraint equation
\ben
\d _{A}I(g,A,\Ab ) &=& {1 \over 2\pi }\int \mbox{Tr} ( \ - \pb g \gi + g\Ab
\gi - \Ab \  )\d A
= 0 \nonumber \\
\d _{\Ab }I(g,A,\Ab ) &=& {1 \over 2\pi }\int \mbox{Tr} ( \  \gi \pp g  +
\gi A g - A \ )\d\Ab = 0 \ ,
\een
describes the coset model at the classical level.

Now, we consider a deformation of the G/H-coset model by adding a primary
field  to the gauged WZW action,
\be
I(g,A,\Ab ,\b ) = I(g,A,\Ab ) + {\b \over 2\pi }\int \mbox{Tr}gT\gi \Tb ,
\ee
where $\b $ is a coupling constant and $T , \Tb$ are elements of the Lie
algebra
$\bf g$ associated with the Lie group G. In the following, we assume that
 $T, \Tb$ belong to the center of the
algebra {\bf h}  so that $
[ ~ \pp + A \ , \ \Tb ~ ] = [~\pb +\Ab \ , \ T ~ ] = 0 $. In such a case, the
equation of motion  once again takes the form of a zero cuvature but
with a spectral parameter $\l$,
\be
[~ \pp + \gi \pp g + \gi A g + \b \l T \ , \ \pb + \Ab +\li \gi \Tb g ~ ] = 0 .
\ee
Since the term $\mbox{Tr} gT\gi \Tb $ is invariant under the vector gauge
transformation, so is the action $I(g,A,\Ab ,\b )$ and
 the constraint equation remains unchanged which we solve
explicitly for $A$ and $\Ab$,
\be
A^{a} = -\sum_{b=1}^{dim \bf h} (M^{-1}(\gi ))^{ab}(\pb g \gi )^{b}
  \ , \ \Ab^{a} = \sum_{b=1}^{dim \bf h}(M^{-1}(g))^{ab}
(\gi \pp g)^{b}
\ee
where $M^{ab}(g) \equiv \d^{ab} -  {1 \over 2}\mbox{Tr}\gi
\T^{a}g\T^{b} $ and $\T^{a}$ are generators of $\bf g$  normalized by
$\mbox{Tr}\T^{a}\T^{b} = 2\d^{ab}$.
By taking the trace of Eq.(5) multiplied with any $\Theta^{a} \in \bf h$,
we may easily see that $A$ and $\Ab$  also satisfy the zero curvature
condition, i.e. $ [ \ \pp + A \ , \ \pb + \Ab \ ] = 0 $
which reflects the vector gauge invariance of the action.
In the following, we solve the zero curvature condition by
 $A= H \pp H^{-1}$ and  $\Ab = H \pb H^{-1}$ for some $H$  and
rewrite $A$ and $\Ab$ in terms of $H$ whenever necessary.
Note that Eq.(5), with $A,  \Ab $ given as in Eq.(6), is precisely  the
integrability condition of a couple of linear equations,
\be
( \pp + \gi \pp g + \gi A g + \b \l T ) \Psi = 0 \  \ ; \ \
( \pb + \Ab + \li \gi \Tb g ) \Psi = 0 \ .
\ee
This linear equation and also the integrability condition  generalize
those of the affine Toda equation which can be solved by applying the inverse
scattering method.[7]
For example, consider a reduction of the GL(N,R)/U(1)-coset
model by setting $g = \mbox{diag}(exp(\phi_{1}), ... , exp(\phi_{N}))$ and
$\Tb_{ij}=T_{ji} $
where $T_{ij} =1 $ if $i-j+1 =0,  $mod(N) and zero otherwise. If we fix the
gauge
by $A=\Ab=0$, then the integrability condition becomes precisely the affine
SL(N,R)-Toda equation.
Therefore, the deformed G/H-coset model constitute a more
general integrable system than the affine Toda theories and as we will show
below,
a similar inverse method can be applied to the deformed G/H-coset model in
obtaining infinitely many conserved currents as well as soliton solutions.

In order to understand the symmetry of deformed coset models,
we transform the linear equation in a different but equivalent form which
makes the equation solvable by iteration.
Let $\Phi \equiv \Psi H^{-1} exp( \l \b Tz)$ so that the linear equation
changes into
\be
\pb \Phi + [~ \Ab \ , \ \Phi ~ ] + \li \gi \Tb g\Phi = 0
\ee
and
\ben
0 &=& \pp \Phi + \Phi \pp H H^{-1} + (\gi \pp g + \gi Ag)\Phi + \l
\b [~ T \ , \ \Phi ~]
\nonumber \\
&=& \pp \Phi + [ ~ A \ , \  \Phi ~ ] +  (\gi \pp g + \gi Ag)_{\bf m}\Phi + \l
\b [~ T \ , \ \Phi ~]
\een
where the subscript $\bf m$ denotes the orthogonal decomposition; $
(\gi \pp g + \gi Ag) =  (\gi \pp g + \gi Ag)_{\bf h} +
(\gi \pp g + \gi Ag)_{\bf m} $ according to the decomposition of the Lie
algebra
${\bf g} = {\bf h} \oplus {\bf m}$ and the constraint equation (3) has been
used.
Eq.(9) may be solved for $\Phi $ iteratively
by assuming $\Phi = \sum_{m = 0}^{\infty }\l^{-m}\Phi_{m}, \ \Phi_{0}=i\b  $
which, combined with Eq.(8), leads to an infinite number of conservation laws,
$\pb J_{m} + \pp \bar{J}_{m} = 0 \ ; \ m \ge 2 $, where
\be
J_{m} \equiv \mbox{Tr}T (\gi \pp g +\gi Ag)_{\bf m}\Phi_{m-1} \ ,
\ \bar{J}_{m} \equiv -  \mbox{Tr}T\gi Tg\Phi_{m-2}  \ .
\ee
For $m=2$ in particular, this becomes the conservation of the stress-energy
tensor $\pp_{\m}T_{\m z}=0$.  Also, note that the integrability condition
Eq.(5) is
invariant under the exchange;
\be
\pp \leftrightarrow \pb \ , \ A \leftrightarrow \Ab \ , \ g \leftrightarrow \gi
\ , \  T
\leftrightarrow \Tb \ , \ \b\l  \leftrightarrow {1\over \l }  \ .
\ee
Repeating Eq.(8) through Eq.(10) with the above exchange, we obtain another
set of infinite number of conserved currents, which together with Eq.(10),
constitute
the conserved currents of the deformed coset model.

Next, we present a systematic way to derive soliton solutions.
First, by setting $\Psi \equiv H \Phi $ we transform the linear equation
into
\be
( \pp + U + \l \b T )\Phi = 0 \  \ ; \  \ (\pb + \li V )\Phi = 0
\ee
where
\be
U \equiv H^{-1}\pp H +H^{-1}\gi \pp gH + H^{-1}\gi AgH \
 , \ V \equiv  H^{-1}\gi \Tb gH \ .
\ee
A trivial solution of Eq.(12) is given by $g = 1 \ , \ A=\Ab = 0\ , \ H = 1 $
and
 $\Phi = \Phi^{o} = exp(-\l \b Tz - \l^{-1} \Tb \zb )$.
 As we show in the following, nontrivial soliton solutions can be  obtained
 from the trivial one by employing the Riemann problem technique with zeros[8].
 Let $\Gamma $ be a closed contour or a contour extending to infinity on the
 complex plane of the parameter $\l $. Consider the matrix function $\psi_{1}
 (z,\zb ,\l)$ which is analytic with $n$ simple zeros $\m_{1}, ... ,\m_{n}$
 inside $\Gamma $ and $\psi_{2}(z,\zb ,\l )$ analytic with $n$ simple
 zeros $\l_{1}, ... , \l_{n}$ outside $\Gamma$. We assume that none of these
zeros lies
on the contour $\Gamma$ and $ \psi_{1} \psi_{2} = 1$ for $\l \ne \m_{i} ,
\l_{i} \ ; \ i = 1,...,n$.
 We normalize $\psi_{1}, \psi_{2}$ by  $\psi_{1}(\infty )=\psi_{2}(\infty ) =
1$.
Differentiating $ \Phi^{o}\Phi^{o -1} = \psi_{1}\psi_{2} = 1 $ with respect to
$z$ and $
\zb $, one can easily see that
\ben
\psi^{-1}_{1}\pp \psi_{1} +\psi^{-1}_{1}\l\b T \psi_{1} &=&
-\pp \psi_{2} \psi_{2}^{-1} + \psi_{2}\l \b T\psi_{2}^{-1}
\nonumber \\
 \psi_{1}^{-1} \pb \psi_{1} + \psi_{1} ^{-1} \li \Tb \psi_{1} &=& -\pb \psi_{2}
\psi_{2}^{-1} - \psi_{2} \li \Tb \psi_{2}^{-1} .
\een
Since $\psi_{1} (\psi_{2})$ is analytic inside (outside) $\Gamma$, we find that
the matrix functions $\tilde{U}$ and $ \tilde{V}$, defined by
\be
\tilde{U} \equiv  -\pp \xi \xi^{-1} + \xi \l \b T\xi ^{-1} - \l \b T
 \ ; \ \tilde{ V}  \equiv  -\l \pb \xi \xi^{-1} +  \xi \Tb \xi^{-1}
\ee
where $\xi = \psi_{1}^{-1}$ or $\xi = \psi_{2}$ depending on the region, become
independent of $\l $. Also, $\tilde{\Phi } \equiv \xi \Phi^{o}$ satisfies the
linear equation;
\be
( \pp + \tilde{U} + \l \b T )\tilde{\Phi } = 0 \ , \ (\pb + \li \tilde{V}
 )\tilde{\Phi } = 0 \ .
\ee
If we make an identification $\tilde{U} = U \ , \ \tilde{V} = V \ , \
\tilde{\Phi } = \Phi $ with $U, V$ as in Eq.(13), $\tilde{U}$ and
$\tilde{V}$ in general provide nontrivial $n$-soliton solutions of deformed
coset
models. In order to prove the existence of  matrix functions
$\psi_{1}$ and $\psi_{2}$ with the properties as stated above, we construct
them explicitly in the following; since $\psi_{1}\psi_{2} = 1$,
 the zeros of $\psi_{1}$ are the poles of $\psi_{2}$ and vice versa.
Thus we consider the ans\"{a}tze;
\be
\psi_{1} = 1+\sum_{l=1}^{n} {B_{l} \over \l -\l_{l}} \ , \
\psi_{2} = 1+\sum_{s=1}^{n} {A_{s} \over \l -\m_{s}}
\ee
where the matrix functions $A_{s}(z,\zb ), B_{l}(z,\zb )$ are to be determined.
The fact that $\psi_{1}\psi_{2}=1$ requires $A_{s}$ and $B_{l}$ to satisfy
algebraic conditions which
may be obtained through the evaluation of residues  of the equation
$\psi_{1}\psi_{2}=1$ at $\l =\l_{l}, \m_{s}$,
\be
A_{s}+\sum_{l=1}^{n}{A_{s}B_{l} \over \m_{s}-\l_{l}} = 0 \ , \
B_{l}+\sum_{s=1}^{n}{A_{s}B_{l} \over \l_{l}-\m_{s}} = 0 .
\ee
Also, the residues at $\l =\l_{l}, \m_{s}$ of Eq.(15) require
\be
(1 + \sum_{s=1}^{n}{A_{s} \over \l_{l}-\m_{s}})(\pp B_{l} +\l_{l}\b T B_{l})
= 0 \ , \
(1 + \sum_{s=1}^{n}{A_{s} \over \l_{l}-\m_{s}})(\pb B_{l}+{1\over \l_{l}}
\Tb B_{l})
= 0
\ee
as well as
\be
(\pp A_{s} -\m_{s}\b A_{s}T)(1+\sum_{l=1}^{n}{B_{l} \over \m_{s}-\l_{l}}) =0
 \ ,  \
 (\pb A_{s}-{1\over \m_{s}}A_{s}\Tb )(1+\sum_{l=1}^{n}{B_{l} \over
\m_{s}-\l_{l}})
=0 .
\ee
In order to solve Eqs.(18)-(20), we assume that
 $(A_{s})_{ij} = m^{s}_{i}n^{s}_{j} \ , \ (B_{l})_{ij} = s^{l}_{i}t^{l}_{j}.$
Then Eq.(18) becomes
\be
s_{i}^{l} +\sum_{k} \sum_{s=1}^{n}{m_{i}^{s}n_{k}^{s}s_{k}^{l}
\over \l_{l}-\m_{s} } = 0 \ , \ n_{j}^{s}-\sum_{k}\sum_{l=1}^{n}{
n_{k}^{s}s_{k}^{l}t_{j}^{l} \over \l_{l} -\m_{s}} = 0
\ee
and Eqs.(19) and (20) become respectively
\be
(\pp + \l_{l}\b T)s^{l} =0 \ , \ (\pb +{1\over \l_{l}}\Tb )s^{l} = 0 \
\mbox{and}
 \ \pp n^{s} - \m_{s}\b n^{s}T = 0 \ , \ \pb n^{s} - {1 \over \m_{s}}n^{s}\Tb
 = 0 \ .
\ee
Note that  $n^{s}$ and $s^{l}$ can be solved in terms of arbitrary constant
  vectors $\bar{n}^{s}$ and $\bar{s}^{l}$,
 \be
n_{i}^{s} = \sum_{j}\bar{n}^{s}_{j} [ \Phi^{o}(\m_{s})^{-1} ] _{ji} \ , \
s^{l}_{i} =\sum_{j} [ \Phi^{o}(\l_{l}) ] _{ij}\bar{s}^{l}_{j} \ ,
\ee
while  $m^{s}$ and $t^{l}$ can be  obtained in terms of $n^{s}$ and $s^{l}$ by
solving the linear algebraic equation (21),
\be
m^{s}_{i} = -\sum_{l=1}^{n}s^{l}_{i}(w^{-1})^{ls} \ , \
t^{l}_{i} = \sum_{s=1}^{n}(w^{-1})^{ls}n_{i}^{s} \ ,
\ee
where $w^{sl} \equiv \sum_{k}n_{k}^{s}s_{k}^{l}/(\l_{l} - \m_{s})$.
Having determined $\psi_{1}$ and $\psi_{2}$ as above,
 we finally obtain $n$-soliton solutions by evaluating  Eq.(15)
 at $\l = \infty $,
\ben
U &=& H^{-1}\pp H +H^{-1}\gi \pp gH + H^{-1}\gi AgH
= \b \sum_{s=1}^{n} ( TB_{s} + A_{s}T)  \nonumber \\
V &=&  H ^{-1}\gi \Tb gH =  -\sum_{s=1}^{n}\pb A_{s} + \Tb .
\een
Of course, instead of  taking the value at $\l = \infty $, we may evaluate
$U, V$ at $\l = 0$ which results in  $U = -\pp \xi (z,\zb ,0)\xi^{-1}(z,\zb ,0)
\ ;
\ V = \xi (z, \zb ,0)\Tb \xi^{-1}(z,\zb ,0)$ leading to the same result.
In general, further restrictions are required for $\psi_{1}$ and $\psi_{2}$
depending on  the specific coset structure. For example, if G and H
are unitary, $U,V$ in the linear equation become anti-hermitian which in turn
impose further restrictions on $\xi$,
\be
\xi^{\dagger }(\l ) = \xi(\l^{*} )^{-1}
\ee
where $\l^{*}$ is the complex conjugation of $\l $.  This  requires that
$\l_{l} = \m^{*}_{l}$ and $A^{\dagger }_{l} = B_{l}$, or $n^{s*}_{i} =
s^{s}_{i}, m^{s*}_{i} = t^{s}_{i}$ so that
$\bar{n}^{s*}_{i} = \bar{s}^{s}_{i}$. Then the matrix $w^{sl}$ becomes
\be
w^{sl}  = (\l_{l} - \lb_{s})^{-1}\sum_{ijk}
\bar{n}^{s}_{i}[\Phi^{o}(\lb_{s})^{-1}]_{ij}
[\Phi^{o}(\l_{l})]_{jk} \bar{n}^{s*}_{k} .
\ee
Beside of unitarity, we may impose other types of restrictions according to
the specific group structure of G as well as
reductions by discrete subgroups of G and H. These cases will be considered
elsewhere.

Having demonstrated a systematic way to construct conserved currents and
soliton
solutions, we take the deformed parafermionic  SU(2)/U(1)-coset model as an
example
and compute explicitly conserved currents and $n$-soliton solutions.
We fix the gauge in such a way that  the group manifold is parametrized by
\be
g = \left( \begin{array}{ccc}
u& ~ & i \sqrt{1-u\ub }\\
{}~           & ~   &  ~ \\
i \sqrt{1-u\ub }& ~ & \ub          \end{array}   \right) .
\ee
We choose the center elements $T = -\Tb = i\s_{3} = \mbox{diag}(i, -i)$.
Then, the connections $A, \Ab$
 are
\be
A = {\ub \pp u -u\pp \ub \over 4(1-u\ub )}\s_{3} \equiv J_{1}\s_{3}
\ , \
\Ab =  {u \pb \ub -\ub \pb u \over 4(1-u\ub )}\s_{3}  \equiv -\bar{J}_{1}\s_{3}
 \ .
\ee
The linear equation (7) now becomes
\be
\pp \Psi + (
{\ub \pp u -u\pp \ub \over 4(1-u\ub )} \s_{3} +
{1 \over \sqrt{1-u\ub }}
  \left( \begin{array}{ccc}
0 & ~ & -i\pp \ub  \\ {}~           & ~   &  ~ \\
 -i\pp u  & ~ &
0          \end{array}   \right)  + i\l \b \s_{3} )\Psi = 0
\ee
and
\be
\pb \Psi + ( {u \pb \ub -\ub \pb u \over 4(1-u\ub )}\s_{3}
-{i \over \l }  \left( \begin{array}{ccc}
2u\ub -1 & ~ & 2i\ub \sqrt{1-u\ub } \\
{}~           & ~   &  ~ \\
 -2iu\sqrt{1-u\ub } & ~ & 1-2u\ub          \end{array}   \right))\Psi = 0
\ee
whose integrability gives rise to the classical equation of motion of $u$
and $\ub$;
\ben
\pp\pb u + {\ub \pb u \pp u \over 1-u\ub } + 4\b u(1-u\ub ) &=& 0 \nonumber \\
\pp\pb \ub + { u \pb \ub \pp \ub  \over 1-u\ub } + 4\b \ub (1-u\ub ) &=& 0
\een
and the U(1)-current conservation law $\pb A -  \pp \Ab = (\pp \bar{J}_{1} +
\pb J_{1})
\s_{3}= 0$ which arises, after
the gauge fixing,  from  the remaining global U(1)-invariance of the action
under $u \rightarrow ue^{i\e }$ and $\ub \rightarrow \ub e^{-i\e }$.
In order to obtain higher conservation laws, we consider the recursive
equation (9) which in the present case is given by
\ben
\pp \Phi_{m} &+& {u\pp \ub -\ub \pp u \over 4(1-u\ub )}[ \ \Phi_{m} \ , \
\s_{3} \ ]  + {1\over \sqrt{1-u\ub}}
 \left( \begin{array}{ccc}
0 & ~ & -i\pp \ub  \\ {}~           & ~   &  ~ \\
 -i\pp u  & ~ & 0
          \end{array}   \right) \Phi_{m} + i\b [ \  \s_{3} \ , \
\Phi_{m+1} \ ] = 0 \nonumber \\
\pb \Phi_{m} &+& {u\pb \ub -\ub \pb u \over 4(1-u\ub )} [ \s_{3} , \Phi_{m} ]
- i  \left( \begin{array}{ccc}
2u\ub -1 & ~ & 2i\ub \sqrt{1-u\ub } \\
{}~           & ~   &  ~ \\
 -2iu\sqrt{1-u\ub } & ~ & 1-2u\ub          \end{array}   \right) \Phi_{m-1} = 0
  .
\een
With $ \Phi_{m} \equiv  \left( \begin{array}{ccc}
p_{m} & ~ & q_{m} \\
{}~           & ~   &  ~ \\
 r_{m} & ~ & s_{m}         \end{array}   \right) $,
 Eq.(33) may be solved iteratively in components;
\ben
p_{m} &=& \int [ dz {i\pp  \ub \over \sqrt{1-u\ub }}r_{m} +
 d\zb (i(2u\ub -1 )p_{m-1} - 2\ub \sqrt{1-u\ub }r_{m-1} )] \nonumber \\
 q_{m} &=& { i \over 2\b }( \pp q_{m-1} +
{\ub \pp u - u\pp \ub \over 2(1-u\ub ) }q_{m-1} -
{i\pp \ub \over \sqrt{1-u\ub }}s_{m-1} ) \nonumber \\
r_{m} &=& - {i \over 2\b }(\pp r_{m-1} -
{\ub \pp u - u\pp \ub \over 2(1-u\ub ) }r_{m-1} -
{i\pp  u \over \sqrt{1-u\ub }}p_{m-1} ) \nonumber \\
s_{m} &=& \int [ dz  {i\pp  u \over \sqrt{1-u\ub }}q_{m} +
d\zb (2u\sqrt{1-u\ub }q_{m-1} + i(1-2u\ub )s_{m-1} )]  \ .
 \een
With $\Phi_{0} = i\b $, the first iterative solution $\Phi_{1}$, for example,
is
\ben
p_{1}=\int [ dz {\pp \ub \pp u \over 2(1- u\ub )} -d\zb \b (2u\ub -1)]
 \ , \ q_{1}=
 {i\pp \ub \over 2\sqrt{1-u\ub }} \nonumber \\
r_{1}= -{i\pp u \over 2\sqrt{1-u\ub }} \ , \
s_{1}=  \int [ -  dz  {\pp \ub \pp u \over 2(1- u\ub )} + d\zb \b (2u\ub -1) ]
 \ .
\een
Higher conservation laws;   $\pb J_{m}  + \pp \bar{J}_{m}  = 0 \  ; \
 m \ge 2$ also result from Eq.(33) with currents $J_{m}, \bar{J}_{m}$ defined
by
\be
J_{m} \equiv {i \pp \ub \over \sqrt{1-u\ub }}r_{m-1} \  \  ; \  \   \bar{J}_{m}
\equiv
2\ub \sqrt{1-u\ub }r_{m-2 } -i(2u\ub -1)p_{m-2} \ .
\ee
For $m = 2$, the conservation law becomes
\be
\pb J_{2}  + \pp \bar{J}_{2}  =
{1\over 2}\pb \left( {\pp u \pp \ub \over 1-u\ub } \right) + \b \pp ( 2u\ub -1)
= 0
\ee
which is the conservation of the stress energy tensor, $\pp_{\m }T_{\m z} = 0$.
 Higher order conserved currents ($m\ge 3$) in general become
non-local.\footnote{
  I would like to thank I.Bakas for raising questions at this point.}
  For example,
\ben
J_{3} &=&{i \over \b}[ -{\pp \ub  \over 2 \sqrt{1-u\ub }}\pp
({\pp u \over \sqrt{1-u\ub }}) + J_{1} J_{2} - J_{2}(\int dz J_{2} -d\zb
\bar{J}_{2} )]
\nonumber \\
\bar{J}_{3} &=& -i \ub \pp u - {i \over \b }\bar{J}_{2}(
\int ( dz J_{2} - d\zb \bar{J}_{2}) .
\een
This however can be made local if we redefine currents
$J^{'}_{3} \equiv J_{3} + i\b ^{-1} J_{2}(\int dz J_{2} - d \zb \bar{J}_{2} )$
and $\bar{J}^{'}_{3} \equiv \bar{J}_{3} + i\b ^{-1} \bar{J}_{2}(\int dz J_{2}
- d\zb \bar{J}_{2} )$ which satisfy $\pp J^{'}_{3} + \pb \bar{J}_{3}^{'} = 0$.

Finally, we calculate explicitly $n$-soliton solutions of  Eq.(32).
Eq.(25) for  the SU(2)/U(1)-case becomes
\be
 \left( \begin{array}{ccc}
2u\ub -1 & ~ &e^{-2ih} 2i\ub \sqrt{1-u\ub } \\
{}~           & ~   &  ~ \\
 -e^{2ih}2iu\sqrt{1-u\ub } & ~ & 1-2u\ub          \end{array}   \right)
= -\sum_{s=1}^{n}i\pb A^{s} +\s_{3}
\ee
where $4ih = \int^{\zb} d\zb^{'}(\ub \pb u - u\pb \ub )(1-u\ub )^{-1}$ and
\ben
A^{s}_{ij} &=& -\sum_{k,m}
\sum_{l=1}^{n}[\Phi^{o}(\l_{l})]_{ik}\bar{n}^{l*}_{k}(w^{-1})^{ls}
\bar{n}^{s}_{m}[\Phi^{o}(\lb_{s})^{-1}]_{mj} \nonumber \\
\Phi^{o}(\l ) &=& exp(-i\b \l \s_{3} z - {1\over i\l }\s_{3}\zb ) \nonumber \\
 w^{sl} &=& (\l_{l} -\lb_{s})^{-1}(\bar{n}_{1}^{s}\bar{n}_{1}^{l*}e^{\L_{sl}} +
\bar{n}_{2}^{s}\bar{n}_{2}^{l*}e^{-\L_{sl}}) \nonumber \\
\L_{sl} &\equiv & i\b (\lb_{s} - \l_{l})z -i({1\over \lb_{s}} - {1\over
\l_{l}})\zb .
\een
If we write $u = re^{i\theta }$,  $n$-soliton solution can be obtained directly
from
Eq.(39) such that
\ben
r^{2} &=& 1-{i \over 2}\sum_{s=1}^{n}\pb A^{s}_{11} \nonumber \\
\theta &=& -{1 \over 4}\int^{\zb } d\zb^{'} \sum_{s=1}^{n}\pb A_{11}^{s} (
\pb ln \sum_{s=1}^{n}\pb A_{12}^{s} - \pb ln  \sum_{s=1}^{n}\pb A_{21}^{s} ) .
\een
In particular, if we choose $n=1$ and write $\l _{1} = \k e^{i\d }$, we obtain
the one soliton
solution,
\ben
r^{2}= {(exp(2\b \k zsin \d + 2\k^{-1} \zb sin \d ) - \eta
exp(-2\b \k zsin \d - 2\k^{-1} \zb sin \d ))^{2} + 4\eta  cos^{2}\d
\over  (exp( 2\b \k zsin \d + 2\k^{-1} \zb sin \d ) +
\eta exp(-2\b \k zsin \d - 2\k^{-1} \zb sin \d ))^{2}}
\een
and
\ben
cos \theta = {exp(2\b \k zsin \d + 2\k^{-1} \zb sin \d ) + \eta cos 2\d
exp(-2\b \k zsin \d - 2\k^{-1} \zb sin \d )  \over \sqrt{(
exp( 2\b \k zsin \d +
2\k^{-1} \zb sin \d )
-\eta exp( -2\b \k zsin \d - 2\k^{-1} \zb sin \d ) )^{2} + 4\eta cos^{2}\d }} \
{}.
\een

In this letter, we have shown that, to each coset G/H,  there exist
a corresponding integrable field theory which arises from a deformation of
coset conformal field theory. At the classical level, these models are shown to
generalize the affine Toda field theories and possess infinite dimensional
symmetries as well as soliton solutions.
The quantum aspect of these models are of great interest and will be considered
elsewhere.
\vglue .2in
{\bf ACKNOWLEDGEMENT}
\vglue .2in
 I am grateful to  Professor  H.J.Shin for many useful discussions and
constructive
 criticism, and to Professors I.Bakas, B.K.Chung, S.Nam, D.Kim and C.Lee for
their help.
This work was supported in part by the program of
Basic Science Research, Ministry of Education,
and by Korea Science and Engineering Foundation.
\vglue .2in

\def\Item{\par\hang\textindent}

{\bf REFERENCES }
\Item {[1]} A.Zamolodchikov, Sov.Phys.JETP Lett. {\bf 46} (1987); Int.J.Mod.
Phys. {\bf A3} (1988) 743.
\Item {[2]} T.Eguchi and S.Yang, Phys.Lett.{\bf B224} (1989) 373.
\Item {[3]} T.Hollowood and P.Mansfield, Phys.Lett.{\bf B226} (1989) 73.
\Item {[4]} D.Karabali, Q-H.Park, H.J.Schnitzer and Z.Yang, Phys.Lett.{\bf
B216} (1989) 307; D.Karabali and H.J.Schnitzer, Nucl.Phys.{\bf B329} (1990)
649;
K.Gawedski and A.Kupiainen, Phys.Lett.{\bf B215}(1988) 119, Nucl.Phys.{\bf
B320} (1989) 625.
\Item {[5]} E.Witten, Commun.Math.Phys.{\bf 92} (1984) 455.
\Item {[6]} I.Bakas, `Conservation laws and geometry of perturbed coset
models',
CERN-TH.7047/93, hep-th/9310122
\Item {[7]} A.V.Mikhailov, Physica {\bf 3D} (1981) 73.
\Item {[8]} V.E.Zakharov, S.V.Manakov, S.P.Novikov and L.P.Pitaievski, Theory
of
Solitons, Moscow, Nauka 1980 (Russian); English transl.:New York, Plenum 1984.
\end{document}